\documentclass{jnmp}

\usepackage{amsmath}

\JNMPnumberwithin{equation}{section}

\newtheorem{theorem}{Theorem}

\theoremstyle{definition}

\def\secref#1{Sec.~\ref{#1}}
\def\eqref#1{(\ref{#1})}
\def\eqrefs#1#2{(\ref{#1}) and~(\ref{#2})}
\def\eqsref#1#2{(\ref{#1})--(\ref{#2})}

\def\EQ{\begin{equation}}
\def\doneEQ{\end{equation}}
\def\EQs{\begin{eqnarray}}
\def\doneEQs{\end{eqnarray}}

\def\eqtext#1{\hbox{\rm{#1}}}

\def\downupindices#1#2{{\mathstrut}^{}_{#1}{\mathstrut}_{}^{#2}}
\def\updownindices#1#2{{\mathstrut}_{}^{#1}{\mathstrut}^{}_{#2}}
\def\mixedindices#1#2{{\mathstrut}^{#1}_{#2}}
\def\downindex#1{{\mathstrut}_{#1}}
\def\upindex#1{{\mathstrut}^{#1}}

\def\u{U}
\def\ut{\u_t}
\def\ux#1{\u_{#1x}}
\def\uflow{\u_{\tau}}

\def\ucc{\bar\u}
\def\utcc{\bar\u_t}
\def\uxcc#1{\bar\u_{#1x}}

\def\v{V}

\def\vx#1{\v_{#1x}}
\def\vvar#1{\v\upindex{#1}}

\def\uvars#1#2{u^{#1}_{#2}}
\def\scu{u}
\def\scut{u_t}
\def\scutt{u_{tt}}
\def\scuttt{u_{ttt}}
\def\scux#1{u_{#1x}}
\def\scuflow{u_{\tau}}

\def\D#1#2{D^{#2}_{#1}}

\def\sprod#1#2{\langle#1,#2\rangle}

\def\Jsp#1{J^{(#1)}}
\def\solJsp#1{R^{(#1)}}
\def\wu{\lambda}
\def\wt{\mu}
\def\wtau{\nu}
\def\wx{\xi}

\def\g#1#2{g\downupindices{#1}{#2}} 
\def\flat#1#2{\eta\downupindices{#1}{#2}} 
\def\ginv#1#2{g\updownindices{#1}{#2}} 

\def\covder#1#2{{}^{#2}\nabla\downindex{#1}}
\def\conx#1#2{\Gamma\updownindices{#1}{#2}}
\def\curv#1#2{R\downupindices{#1}{#2}}
\def\scurv{\chi}

\def\id#1#2{\delta\downupindices{#1}{#2}}

\def\der#1{\partial\downindex{#1}}
\def\nder#1#2{\partial\mixedindices{#2}{#1}}

\def\deru#1{\frac{\partial}{\partial \uvars{#1}{}}}

\def\algconx#1#2{{}^\alg{#1}\omega\downindex{#2}}
\def\e#1#2{e\mixedindices{#1}{#2}}
\def\algtconx#1#2{{}^\alg{#1}\omega_t\upindex{#2}}
\def\algxconx#1#2{{}^\alg{#1}\omega_x\upindex{#2}}
\def\te#1{e_t\upindex{#1}}
\def\xe#1{e_x\upindex{#1}}

\def\quot{p}
\def\quotvec#1{\quot\upindex{#1}}
\def\matrh{{\bf h}}

\def\matru{{\mathcal U}}

\def\alg#1{{\mathfrak #1}}
\def\T{{\rm T}}
\def\I{{\mathbb I}}
\def\zero{{\bf 0}}

\def\R#1{{\mathbb R}^{#1}}
\def\C{{\mathbb C}}

\def\d{{\rm d}}

\def\i{{\rm i}}

\def\nnwp/{nonnegatively-weighted polynomial}

\def\hsymm/{higher symmetry}
\def\hsymms/{higher symmetries}
\def\lhs/{l.h.s.\!}
\def\rhs/{r.h.s.\!}
\def\ie/{i.e.}

\begin{document}

%  Headings
%
\renewcommand{\evenhead}{S.C. Anco and T. Wolf}
\renewcommand{\oddhead}{Some symmetry classifications 
of hyperbolic vector evolution equations}

%  Titlepage
%
\thispagestyle{empty}

\FirstPageHead{12}{1}{2005}
{\pageref{firstpage}--\pageref{lastpage}}{{\bf \tiny
{Birthday Issue}}}

\copyrightnote{2005}{S C Anco and T Wolf}

\Name{Some Symmetry Classifications of 
Hyperbolic Vector Evolution Equations}

\label{firstpage}

\Author{Stephen C. ANCO~$^\dag$ and Thomas WOLF~$^\ddag$}

\Address{
Department of Mathematics, Brock University, Canada L2S 3A1
\\
~~E-mail: $^\dag$sanco@brocku.ca ,$^\ddag$twolf@brocku.ca}

\Date{This article is part of the special issue published
in honour of Francesco Calogero on the occasion of his 70th birthday}

\begin{abstract}
\noindent
Motivated by recent work on integrable flows of curves
and 1+1 dimensional sigma models,
several $O(N)$-invariant classes of hyperbolic equations
$\ux{t} =f(\u,\ut,\ux{})$
for an $N$-component vector $\u(t,x)$ are considered. 
In each class 
we find all scaling-homogeneous equations 
admitting a \hsymm/ of least possible scaling weight. 
Sigma model interpretations of these equations are presented. 
\end{abstract}

\section{Introduction}
\label{intro}
\resetfootnoterule

Integrability theory of nonlinear partial differential equations (PDEs)
has many different aspects which have developed over the years ---
B\"acklund transformations, 
Lax pairs, 
inverse scattering transform and soliton solutions 
(Calogero's ``S-integrability''), 
linearizing transformations (``C-integrability''), 
reduction to Painlev\'e transcendents, 
hierarchies of \hsymms/ and conservation laws, 
master symmetries, 
recursion (Nijenhuis) operators,
bi-Hamiltonian structures,
and others. 
To date the most computationally direct and effective
test of integrability has proved to be the condition that
a PDE system should possess sufficiently many \hsymms/ 
\cite{IbragimovShabat,Fokas}
(see also \cite{symmintegr1,symmintegr2,symmintegr3,symmintegr4}). 
In particular, 
for all currently known examples of nonlinear scalar PDEs, 
the existence of one \hsymm/ implies the existence of 
infinitely many, \ie/ a symmetry hierarchy. 

A rigorous proof of the observation ``one symmetry implies infinitely many'' 
was established in 
\cite{SandersWang-integrability,BeukersSandersWang}
for a wide class of semilinear scalar evolutionary PDEs
\EQ
\label{scalarutPDE}
\scut=\scux{n} + f(\scu,\scux{},\ldots,\scux{n-1}) ,\quad n>1,
\doneEQ
under a homogeneity restriction 
with respect to a scaling symmetry on $t,x,\scu$
(so-called $\wu$-homogeneous form). 
The proof uses the symbolic method of Gel'fand-Dikii, 
combined with computer algebra computations. 
As a main result, it was shown that 
any such symmetry-integrable scalar polynomial PDE 
with positive scaling weight ($\lambda>0$) for $\scu$
belongs to one of the well known hierarchies 
\cite{integrablehierarchies1,integrablehierarchies2}:
$n=2$ Burger's ($\wu=1$); 
$n=3$ Korteweg de Vries (KdV) ($\wu=2$), 
modified Korteweg de Vries (mKdV) ($\wu=1$), 
Ibragimov-Shabat (IS) ($\wu=1/2$);
$n=5$ Kaup-Kupershmidt (KK) ($\wu=2$), 
Sawada-Kotera (SK) ($\wu=2$), 
Kupershmidt (K) ($\wu=1$), 
potential Kaup-Kupershmidt ($\wu=1$), 
and potential Sawada-Kotera ($\wu=1$). 
The main evolutionary PDEs of semilinear form \eqref{scalarutPDE}
not covered by this classification are 
the hierarchy of the nonlinear Schr\"odinger (NLS) equation 
and its various derivative-type generalizations 
\cite{integrablehierarchies1}, 
in which $\scu$ is complex-valued
so its real and imaginary parts satisfy 
a pair of coupled scalar evolutionary PDEs. 
There has been much work and interest in finding multi-component 
generalizations of all these hierarchies,
particularly where $\scu$ is replaced by 
a vector $\u$ or scalar-vector pair $(\scu,\u)$ or matrix $\matru$. 
Recent symmetry-integrability classification results in this direction 
have been obtained by 
extensive computer algebra computations 
\cite{OlverSokolov,SokolovWolf,TsuchidaWolf02}. 
These classifications test for the existence of a \hsymm/,
\footnote{ 
It should be noted that 
there are special classes of Bakirov-type 
multicomponent evolutionary PDE systems that, surprisingly,
admit only finitely many \hsymms/
\cite{counterexamples}.
However, all these systems are of a triangular form in which one PDE is
a decoupled linear evolution equation. 
In contrast, 
the classifications 
in \cite{OlverSokolov,SokolovWolf,TsuchidaWolf02}
in addition to our results 
deal with fully coupled, nonlinear multicomponent PDE systems,
describing a single vector or matrix nonlinear evolution equation 
for $\u$ or $\matru$,
or a non-triangular scalar-vector pair of nonlinear evolution equations
for $(\scu,\u)$.
}
in analogy with the scalar case. 

Comparatively less work has been devoted to investigating 
symmetry-integrability of hyperbolic PDE systems 
\cite{ZhiberShabat}, 
apart from the sine-Gordon (SG) equation and its scalar variants
related to the $S^2$ sigma model,
which both have the wave equation form 
$\scux{t} = f(\scu,\scut,\scux{})$
and possess zero scaling-weight for $\scu$. 

In this paper we present some symmetry-integrability classifications of
hyperbolic vector PDEs. 
The classifications are motivated by the following three types of 
examples of scalar wave equations:
\vskip 0pt
$(1)$ the SG equation $\scux{t}=\sin\scu$, 
which is distinguished by scaling weight $\lambda=0$ for $\scu$; 
\vskip 0pt
$(2a)$ the variant SG equation $\scux{t} = \scu\sqrt{1-\scut^2}$
coming from the transformation $\scu\rightarrow \sin^{-1}(\scut)$,
which has scaling weight $\lambda=1$ for $\scu$ 
and shares the same \hsymms/ as the mKdV hierarchy;
\vskip 0pt
$(2b)$ the complex-valued version $\scux{t} = \scu\sqrt{1-|\scut|^2}$,
which shares the \hsymms/ of the NLS hierarchy; 
\vskip 0pt
$(3)$ the mKdV equation in potential form 
$\scux{t} = \scux{4} + \scux{}^2\scux{2}$
and the NLS equation in potential form 
$\i\scux{t} = \scux{3} + |\scux{}|^2\scux{}$. 

Our results give interesting vector counterparts of 
the first two types of wave equations. 
We show that these types of hyperbolic vector equations all have 
a natural geometric origin in terms of 
integrable nonlinear sigma models
and integrable inverse flows of curves 
in Riemannian manifolds. 
For the third type of wave equation, 
we find a null result that no vector counterparts exist.

\section{ Classification Results }
\label{results}

For an $N$-component vector $\u(t,x)$, 
with $N$ arbitrary, 
consider an $O(N)$-invariant vector PDE of generalized evolutionary form
\EQ
\label{vectorutmxPDE}
\ux{t\,m} = f(\u,\ux{},\ldots,\ux{n},\ut,\ldots,\ux{t\,m-1}) ,\quad
n,m \geq 0
\doneEQ
namely $f$ is a sum of terms proportional to its vector arguments
with scalar coefficients depending on dot products 
$\sprod{\cdot}{\cdot}$ of these vectors. 
Such a vector PDE is strictly {\it evolutionary} if $m=0,n\geq 1$, 
or is {\it hyperbolic} if $m=1,n\leq 1$; 
we call it a {\it wave equation} if $m=1,n\geq 0$. 
In all cases 
it is {\it semilinear} whenever $n\leq m$ 
or if $f$ is linear in $\ux{n}$ when $n>m$. 
We say it is of {\it minimal differential order} $(m,n)$ 
provided $f$ is not a total derivative with respect to $x$. 
The order of the vector PDE \eqref{vectorutmxPDE} 
in the ordinary sense is $\max(n,m+1)$. 

If we view a vector PDE \eqref{vectorutmxPDE} as formally defining 
the generator of a flow $\D{t}{}\D{x}{m}$ in the jet space 
$\Jsp{\infty}=(t,x,\u,\ut,\ux{},\ldots,\ux{kt\,l},\ldots)$,
then a {\it \hsymm/}
on the solution jet space 
$\solJsp{\infty} \subset \Jsp{\infty}$
(\ie/ modulo equation \eqref{vectorutmxPDE} and differential consequences)
will be a commuting flow $\D{\tau}{}$
whose generator is of the form 
\EQ
\label{vectorusymm}
\uflow = g(\u,\ux{},\ldots,\ux{r},\ut,\ldots,\ux{t\,m-1})
\doneEQ
for some order $r>\max(n,m+1)$,
where $g$ is a sum of terms proportional to its vector arguments
with scalar coefficients depending on dot products 
$\sprod{\cdot}{\cdot}$ of these vectors. 
Thus the \hsymm/ condition is 
\EQ
[\D{\tau}{},\ \D{t}{}\D{x}{m}] =0 
\quad\eqtext{ on $\solJsp{\infty}$ .}
\doneEQ

A flow \eqref{vectorutmxPDE} 
is said to be $\wu$-homogeneous of weight $(\wu,\wt)$
if it admits a scaling symmetry group 
(with parameter $\epsilon \in \R{}$)
\EQ
\label{scaling}
(t,x,\u) \rightarrow 
(e^{\wt\epsilon}t,e^{\epsilon}x,e^{-\wu\epsilon}\u) . 
\doneEQ
Note, with this convention for scaling weights,
the order of a semilinear flow \eqref{vectorutmxPDE} is equal to the weight 
$\wt$ in the evolutionary case
and $\wt+1$ in the hyperbolic case. 
Without loss of generality,
\hsymms/ \eqref{vectorusymm} for such flows 
can be assumed to be homogeneous under 
the same scaling symmetry group \eqref{scaling} 
(since any symmetry necessarily decomposes into a sum of terms
with homogeneous forms which themselves will define symmetries). 
Correspondingly,
we will refer to \hsymms/ \eqref{vectorusymm}
by their weight as defined through the scaling 
\EQ
\tau \rightarrow 
e^{\wtau\epsilon}\tau
\doneEQ
induced by the scaling symmetry group \eqref{scaling}. 
Note this weight $\wtau$ is equal to the order $r$
precisely when a \hsymm/ is of semilinear form, \ie/ 
linear in $\ux{r}$. 

Throughout, 
we will assume a partial polynomial form 
for $\wu$-homogeneous vector PDEs \eqref{vectorutmxPDE}
and \hsymms/ \eqref{vectorusymm}. 
More specifically, as explained in \secref{computation}, 
$f$ and $g$ will be linear in their vector arguments 
and polynomial in all nonnegative-weight dot products of these vectors,
with undetermined coefficients that either 
are functions of any zero-weight dot products or otherwise are constants. 
We refer to such a form as the {\it \nnwp/ class} of 
vector PDEs and \hsymms/.
Hereafter we restrict attention to scaling weights 
$\wu=0,\frac{1}{2},1,2$
in analogy with the classes of symmetry-integrable scalar evolutionary PDEs
summarized in \secref{intro}. 

In the case of evolutionary vector PDEs,
the following classification was obtained in \cite{SokolovWolf}. 
Up to scalings of $t,x,\u$,
the only vector counterparts of 
$\wu$-homogeneous evolutionary scalar polynomial PDEs \eqref{scalarutPDE}
with scaling weights $\wu=\frac{1}{2},1,2$ 
possessing a homogeneous \hsymm/ of least possible weight
are given by, in the case of real-valued $\u$:
\vskip0pt
$\bullet$
vector mKdV equations $\wu=1,\wt=3$, 
\EQs
&&
\ut = \ux{3} + \sprod{\u}{\u}\ux{} , 
\label{mkdveqI}\\
&&
\ut = \ux{3} + \sprod{\u}{\u}\ux{} + \sprod{\u}{\ux{}}\u{} ; 
\label{mkdveqII}
\doneEQs
\vskip0pt
$\bullet$
vector IS equation $\wu=\frac{1}{2},\wt=3$, 
\EQ
\ut = 
\ux{3} + 3\sprod{\u}{\u}\ux{2} + 6\sprod{\u}{\ux{}}\u
+  3\sprod{\u}{\u}{}^2 \ux{} + 3\sprod{\ux{}}{\ux{}}\u ;
\label{iseq}
\doneEQ
and in the case of complex-valued $\u$:
\vskip0pt
$\bullet$
vector NLS equations $\wu=1,\wt=2$, 
\EQs
&&
\i\ut = 
\ux{2} \pm \sprod{\u}{\ucc}\u , 
\label{nlseqI}\\
&&
\i\ut = 
\ux{2} \pm 2\sprod{\u}{\ucc}\u \mp \sprod{\u}{\u}\ucc ; 
\label{nlseqII}
\doneEQs
\vskip0pt
$\bullet$
vector derivative-NLS equations $\wu=\frac{1}{2},\wt=2$, 
\EQs
\i\ut &=&
\ux{2} + a_1 \sprod{\u}{\ucc} \ux{} + a_2 \sprod{\u}{\u} \uxcc{} 
+ a_3 \sprod{\u}{\uxcc{}} \u + a_4 \sprod{\ucc}{\ux{}} \u 
+ a_5 \sprod{\u}{\ux{}} \ucc 
\nonumber\\&&
+a_6 \sprod{\u}{\u} \sprod{\ucc}{\ucc} \u
+a_7 \sprod{\u}{\u} \sprod{\u}{\ucc} \ucc
+a_8 \sprod{\u}{\ucc}{}^2 \u
\label{dernlseq}
\doneEQs
whose coefficients $a_1,\ldots,a_8$ depend on two constant parameters,
falling into 6 classes, 
as listed in \cite{SokolovWolf}.
We remark that this classification does not consider
the complex-valued case for $\wt=3$ or $\wt=5$,
\ie/ vector analogs of complexly-coupled 
mKdV, IS, K, KK, SK, and potential KK, SK equations,
other than ones contained in the NLS hierarchy. 
Hereafter, in the case of complexed-valued $\u$,
we allow $t\rightarrow \i t$, $\tau\rightarrow \i \tau$, 
and impose invariance under the phase symmetry 
$\u \rightarrow e^{\i\epsilon}\u$ ($\epsilon\in\R{}$),
as displayed by NLS type systems. 

\subsection{ Vector-potential wave equations }

We now proceed to discuss symmetry-integrability classifications
of vector $\ux{t}$ equations. 
Our results in theorems~\ref{potmkdvclass}--\ref{mkdvclass}
are established by computer algebra computations, 
as outlined in detail in \secref{computation}. 
In each case we explicitly determine all equations 
admitting a higher symmetry of least weight. 

The first class we will consider consists of potential equations
related to the integrable evolutionary vector equations 
\eqsref{mkdveqI}{dernlseq}
via $\u=\vx{}$, with scaling weight $\wu_\v= \wu-1$. 
Note that the only ones whose potential form 
has non-negative scaling weight $\wu_\v$
are the vector mKdV equations and NLS equations with $\wu_\v=0$. 
A natural question is whether they still possess \hsymms/
in potential form. 
We will settle a slightly more general classification problem
by considering $\wu_\v$-homogeneous vector potential wave equations 
\EQ
\label{vectorpotutPDE}
\vx{t} = f(\v,\vx{},\ldots,\vx{n}) ,\quad
n\geq 0
\doneEQ
where $f$ is a sum of terms proportional to the vectors $\vx{l}$,
$0\leq l\leq n$, 
with coefficients depending on the dot products 
$\sprod{\vx{j}}{\vx{k}}$. 

\begin{theorem}
\label{potmkdvclass}
In the \nnwp/ class 
for the potential-mKdV weight $(\wu_\v,\wt)= (0,3)$, 
every $\wu_\v$-homogeneous vector wave equation \eqref{vectorpotutPDE}
that possesses a homogeneous \hsymm/ of weight $\wtau=5$
containing no $t$ derivatives 
is linear.
The same result is true for the potential-NLS weight $(\wu_\v,\wt)= (0,2)$
with $\wtau=3$. 
\end{theorem}

In contrast to this null result,
there are many nonlinear {\it scalar} $\scux{t}$ equations 
possessing a \hsymm/:
for example, 
the potential-mKdV equation 
$\scux{t}= \scux{4} + \scu^2\scux{}$,
and the 
3rd-order Burger's equation in potential form
$\scux{t}= 
\scux{4} + \scux{}\scux{3} + \scux{2}{}^2+\frac{1}{3} \scux{}{}^2\scux{2}$. 

\subsection{ Vector sine-Gordon equations }

The next class of vector wave equations we investigate
will be hyperbolic vector variants of the SG equation 
$\scux{t} =\sin\scu$, 
distinguished by the scaling symmetry 
$(t,x,\scu) \rightarrow (e^{-\epsilon}t,e^{\epsilon}x,\scu)$. 
Although the transcendental form of nonlinearity in the SG equation
precludes any obvious vector analog,
we note there are other integrable hyperbolic scalar equations
with the same scaling symmetry 
yet compatible with a vector nature for $\scu$,
such as the equation 
$\scux{t} =-\scu(1-\scu^2)^{-1} \scut\scux{}$
which is related to the scalar wave equation 
$\scux{t}=0$ by the transformation $\scu \rightarrow \sin\scu$. 
We now classify 
analogous hyperbolic vector equations 
possessing \hsymms/ with the same weight as the ones in both 
the SG hierarchy 
and the hierarchy connected with the scalar wave equation. 

\begin{theorem}
\label{sgclass}
In the \nnwp/ class
for the SG weight $(\wu,\wt)=(0,-1)$,
every $\wu$-homogeneous nonlinear hyperbolic vector evolution equation 
\eqref{vectorutmxPDE}
that possesses a \hsymm/ of weight $\wtau=3$ 
is given by (up to scalings of $t,x,\u$)
\EQs
&&
\ux{t} = -\sprod{\ut}{\ux{}}(\sprod{\u}{\u}+\alpha)^{-1} \u ,\quad
\alpha=-1,0,1
\label{vectorSGeq}\\
&&
\uflow = \sprod{\u}{\ux{}}^2 \ux{} . 
\doneEQs
This result extends at least to weights $\wtau=4,5,6,7$, 
and in particular 
\EQs
&&
\uflow = 
( \sprod{\u}{\ux{3}}+3 \sprod{\ux{}}{\ux{2}} ) \sprod{\u}{\ux{}} \ux{}
+ ( \sprod{\u}{\ux{2}}+\sprod{\ux{}}{\ux{}} )^2 \ux{}
\notag\\&&\qquad\qquad
+\alpha( \sprod{\ux{}}{\ux{3}} + \sprod{\ux{2}}{\ux{2}} ) 
\ux{} 
\label{vectorSGsymm}
\doneEQs
is an admitted \hsymm/ of order $3$ for weight $\wtau=5$. 
\end{theorem}
We will give a geometrical interpretation 
for this wave equation 
in terms of a $N$-dimensional sigma model
based on a paraboloid Riemannian space 
in \secref{sigmamodels},
where we also discuss the question of its integrability. 

\subsection{ Vector hyperbolic flow equations }

The final class of vector wave equations we will consider
is motivated by the SG variant equation
$\scux{t} = \scu\sqrt{1-\scut^2}$
whose associated flow commutes with the mKdV hierarchy of \hsymm/ flows. 
With respect to this hierarchy,
this wave equation represents a $-1$ flow
as it is mapped into the trivial flow $\scuflow=0$
by the mKdV recursion operator. 
We recall that the recursion operator generates 
the mKdV hierarchy starting from the $x$-translation symmetry 
$\scuflow=\scux{}$,
referred to as the $0$ flow for the hierarchy;
the $+1$ flow is the mKdV equation itself 
$\scuflow = \scux{3} + \frac{3}{2}\scu^2\scux{}$,
while its \hsymms/ define the $+2$ flow and so on. 
We now systematically look for hyperbolic vector equations
that describe $-1$ flows for the hierarchies of
each integrable vector evolutionary equation \eqsref{mkdveqI}{dernlseq}.
In general, the $-1$ flow corresponding to a $\wu$-homogeneous $+1$ flow
in the vector case will have weight $(\wu,-1)$,
as we explain next. 

Consider first the case of the mKdV and IS hierarchies
where the $0$ flow is $\uflow=\ux{}$ 
with scaling weight $\wtau_0$ equal to $1$. 
Since the scaling weight of the $+1$ flow is $\wtau_1=3$,
the weight difference between successive flows in the hierarchy 
is $w=\wtau_1-\wtau_0=2$. 
Hence if the $-1$ flow is regarded as a nonlocal evolutionary equation
$\ut= \D{x}{-1}f(\u,\ut,\ux{})$
then it will have scaling weight $\wt=\wtau_0-w=2\wtau_0-\wtau_1=-1$. 
The remaining case is the NLS and derivative-NLS hierarchies
where the $0$ flow is $\uflow=\i\u$ 
with scaling weight $\wtau_0$ equal to $0$. 
The $+1$ flow is then $\uflow=\ux{}$,
and the (derivative-) NLS equation itself corresponds to the $+2$ flow,
with scaling weight $\wtau_2=2$. 
The weight difference between flows in this case is $w=\wtau_1-\wtau_0=1$,
and so the $-1$ flow will then have scaling weight 
$\wt=0-w=2\wtau_0-\wtau_1=-1$.
Therefore in all cases 
the weight $\wt=-1$ of $t$ for the $-1$ flow is simply the negative of
the weight of $x$. 
A similar argument clearly applies to the well-known scalar hierarchies. 

We now state classifications that 
determine all $-1$ flows in the vector case.

\begin{theorem} (mKdV and NLS $-1$ flows)
\label{mkdvclass}
For $\wu$-homogeneous nonlinear hyperbolic vector evolution equations 
\eqref{vectorutmxPDE}
with weight $(\wu,\wt)= (1,-1)$
in the \nnwp/ class,
the ones that possess a \hsymm/ of mKdV weight $\wtau=3$ 
are given by (up to scalings of $t,x,\u$)
\EQs
&&
\ux{t} 
= \pm \sqrt{ 1-\alpha \sprod{\ut}{\ut} } \u ,
\quad 
\alpha = -1,1
\label{invmkdvIflow}\\
&&
\uflow 
= \ux{3} + \frac{3}{2}\alpha \sprod{\u}{\u} \ux{} , 
\label{invmkdvIsymm}
\doneEQs
and
\EQs
&&
\ux{t} 
= \sprod{\u}{\ut}\sprod{\ut}{\ut}{}^{-1} (1 + A ) \ut 
- |\alpha| (1 - A ) \u , 
\quad
\alpha = -1,0,1
\label{invmkdvIIflow}\\
&&
\uflow 
= \ux{3} + 3\alpha \sprod{\u}{\u} \ux{} + 3\alpha \sprod{\u}{\ux{}} \u , 
\label{invmkdvIIsymm}\\
&& \eqtext{\em where } 
A= \pm\sqrt{ 1-\alpha \sprod{\ut}{\ut} } , 
\doneEQs
while the ones that possess a \hsymm/ of NLS weight $\wtau=2$
are given by (up to scalings of $t,x,\u$)
\EQs
&&
\ux{t} 
= \sprod{\u}{\utcc}\sprod{\ut}{\utcc}{}^{-1} (\alpha + B) \ut
- (\alpha - B) \u ,
\quad
\alpha =-1,1
\label{invnlsIflow}\\
&&
\i\uflow 
= \ux{2} +\frac{1}{4} \sprod{\u}{\ucc} \u
\label{invnlsIsymm}\\
&& \eqtext{\em where }
B = \pm\sqrt{ \alpha^2 -\frac{1}{2}\sprod{\ut}{\utcc} } , 
\doneEQs
and 
\EQs
&&
\ux{t} 
= \sqrt{ C_2 +\sprod{\ut}{\utcc} +\alpha }\; C_1^{-1}\;
\left( \sprod{\u}{\ut} \utcc - \sprod{\u}{\utcc} \ut \right)
%\notag\\&&\qquad\qquad
+ \sqrt{ C_2 -\sprod{\ut}{\utcc} -\alpha }\; \u ,\qquad\quad
\label{invnlsIIflow}\\
&&
\i\uflow 
= \ux{2} +\frac{1}{2} \sprod{\u}{\ucc} \u
- \frac{1}{4} \sprod{\u}{\u} \ucc ,\quad
\alpha =-1,0,1 
\label{invnlsIIsymm}\\
&& \eqtext{\em where }
C_1 
= \sqrt{ \sprod{\ut}{\ut}\sprod{\utcc}{\utcc} - \sprod{\ut}{\utcc}{}^2 } ,
\label{invnlsIIcoeffs}\\
&& 
C_2 
= \pm \sqrt{ \alpha^2 +2\alpha \sprod{\ut}{\utcc} 
+ \sprod{\ut}{\ut}\sprod{\utcc}{\utcc} } . 
\notag
\doneEQs
\end{theorem}
All four of these $-1$ flow equations
are expected to be integrable, 
as they possess a $+1$ flow of the vector mKdV hierarchies
or a $+2$ flow of the vector NLS hierarchies
(up to scaling of $t,x,\u,\tau$). 
We will interpret the first two of them 
geometrically in terms of homogeneous $N+1$-dimensional gauged sigma models 
in \secref{sigmamodels}. 

Finally, we mention that 
a classification of IS and derivative-NLS $-1$ flows
analogous to theorem~\ref{mkdvclass}
in the \nnwp/ class with weight $(\wu,\wt)= (\frac{1}{2},-1)$
yields a null result. 
However, such flows do exist in a wider class, 
which we will present elsewhere.

\section{Computational Aspects}
\label{computation}

Here we describe the algorithmic and implementation issues
involved in the computations used to prove the classification theorems
in \secref{results}.
These computations divide into three parts:
\vskip0pt
(1) making an ansatz for the PDEs and symmetries 
with specified homogeneity weights; 
\vskip0pt
(2) computing the symmetry conditions;
\vskip0pt
(3) solving the system of symmetry conditions. 
\vskip0pt\noindent
The first two tasks are handled by the program {\sc SVSym}
(Scalar-Vector-Symmetry) 
\footnote{{\sc SVSym} is a completely new program, 
more general than the earlier program used 
for the classification of evolutionary PDEs 
in \cite{SokolovWolf,TsuchidaWolf02},
which was limited to positive weights and a strictly polynomial ansatz.}
and the third computation is performed by the
package {\sc Crack} described in \cite{Wolf02}. 

The class of PDE systems for investigation is allowed to involve 
any number of scalar functions $u$ and vector functions $U$ of $t,x$,
which we denote collectively as $\phi^i$. 
An essential limitation currently in the program is that 
the left-hand side (\lhs/) of each PDE must consist of 
a single derivative $\phi^i_{pt\,qx}$ 
sharing the same order $p,q \geq 0$ for all PDEs in the system. 

It is assumed that 
all derivatives occurring on the right-hand side (\rhs/) of
the PDEs and the symmetries 
are not equal to any \lhs/ derivatives or their differential consequences. 
In addition, 
\rhs/ derivatives must have a lower priority with respect to 
some total ordering $>_T$ of derivatives 
than all \lhs/ derivatives,
to avoid infinite substitution loops
in computing the symmetry condition. 
For example, 
the equation 
$\scux{t}=\scutt+\scux{x}$
would lead to an infinite loop of substitutions 
for differential consequences of $\scux{t}$, as seen by 
$\scux{tx} \rightarrow \scux{tt}+\scux{xx}$, 
$\scux{tt} \rightarrow \scuttt+\scux{tx}$. 
This problem is avoided if we adopt (without loss of generality) 
the lexicographical ordering $t >_T x$. 
Then because we have $\scutt >_T \scux{t}$, 
this equation would be handled by bringing the highest priority derivatives
to the \lhs/, 
$\scutt= \scux{x} - \scux{t}$, 
which no longer leads to any infinite substitution loops. 

\subsection{Specifying an Ansatz}

Many integrable systems are $\lambda$-homogeneous 
with respect to
multiple homogeneity weights. 
For example, denoting the weight for $x$ as $\xi$, 
the vector NLS system \cite{SokolovWolf}
\EQs
U_t =   U_{xx} + \sprod{U}{V} U ,\quad
V_t = - V_{xx} - \sprod{U}{V} V
\notag
\doneEQs
and its \hsymm/ hierarchy 
is $\lambda$-homogeneous 
with weights $\mu,\xi,\lambda_U,\lambda_V$ equal to 
$2,1,1,1$, as well as $0,0,1,-1$. 
The latter weights have been used as an extra filter 
in our investigation of PDEs in the complex-valued case. 
To accommodate multiple weightings, 
the program {\sc SVSym} allows the specification of a list
of weight sets $\{\mu,\xi,\nu,\lambda_i\}$.

The program places two a priori restrictions on possible weights
to ensure that the number of possible homogeneous terms 
allowed to appear in the PDEs and symmetries is necessarily finite. 
A first restriction is that 
$x$ cannot have zero weight in all weight sets 
as this would permit an unlimited differential order
of $x$- derivatives 
(note the differential order of $t$- derivatives 
is always finite due to the ordering $t >_T x$
combined with the fixed number $p$ of $t$- derivatives on the \lhs/). 
A second restriction is that 
there must be a weight set $\{\wt^*,\wx^*,\wtau^*,\wu_i^*\}$ 
for which the total weight of the \lhs/ of 
the PDEs $\phi^i_{pt\,qx}$ and symmetries $\phi^i_{\tau}$ 
is non-negative, 
namely 
$w^{\rm PDE}_i:= \wu_i^* + p\wt^* + q\wx^* \geq 0$ 
(with $\wx^* \neq 0$), 
and 
$w^{\rm sym}_i := \wu_i^* + \wtau^* \geq 0$, 
as then the \rhs/ of the PDEs and symmetries
can be generated by the following finite ansatz:
\EQ
u^i_{pt\,qx}  = f^i ,\quad
U^i_{pt\,qx} = \sum_{k,J} F^{ikJ} U^k_J ,\quad
\eqtext{ and }\quad
u^i_\tau  =  g^i ,\quad
U^i_\tau  = \sum_{k,J} G^{ikJ} U^k_J . 
\label{rhseqnsym}
\doneEQ
Here $\scriptstyle J$ is a jet index, denoting 
partial derivatives including those of zeroth order
(only $t$- and $x$-derivatives of $u^k,U^k$ are used
whose total ordering with respect to $t>_T x$ is 
lower than that of the \lhs/ of the PDEs). 
In \eqref{rhseqnsym}, 
$U^k_J$ has $^*$-weight $\geq 0$; 
$f^i, g^i, F^{ikJ}, G^{ikJ}$ are the most general polynomials
(of appropriate weight) 
that can be built from 
the following scalar polynomial variables, 
all of which are constructed to have positive $^*$-weight: 
$u^k_J$, products $u^{k_1}_{J_1}u^{k_2}_{J_2}$
where one of the two factors has a negative $^*$-weight, 
and dot products $\sprod{U^{k_1}_{J_1}}{U^{k_2}_{J_2}}$ 
where at most one of the two factors may have a negative $^*$-weight. 
The undetermined coefficients in all polynomials 
are arbitrary functions of similarly constructed scalar variables
$u^k_J$, $u^{k_1}_{J_1}u^{k_2}_{J_2}$, $\sprod{U^{k_1}_{J_1}}{U^{k_2}_{J_2}}$
whose weight is zero in all weight sets.
All weight sets other than the $^*$-weight set 
are used as an extra filter on the generated terms.
(The restriction of using polynomial variables built from at most
two scalar or vector factors could obviously be generalized to allow
more factors such that the total weight of the product still satisfies
the previous conditions.)

It is important to emphasize that 
the ansatz \eqref{rhseqnsym} for the PDEs and symmetries 
need not be strictly polynomial 
if $\wt^*<0$ or $\wx^*<0$ or any $\wu_i^*\leq 0$.

A few flags give convenient means to cut down the generality of the ansatz, 
for example by discarding $t$-derivatives or imposing constant coefficients 
in the PDEs or symmetries or both. 
Extra conditions or inequalities on coefficients 
can be added easily. 

The following examples show the ansatz generated for 
the classifications in theorems~\ref{sgclass} and~\ref{mkdvclass}. 
The vector SG ansatz 
with $\wt=-1,\wtau=3, \wu_U=0$ is:
\begin{eqnarray*}
\ux{t}\! &\! =\! &\! 
a_1 \u ,
\\ 
\uflow &\! =\! &\! 
b_{1} \ux{3} 
+ b_{2} \ux{2} \sprod{\ux{}}{\u}
+ b_{3} \ux{} \sprod{\ux{2}}{\u} 
+ b_{4} \ux{} \sprod{\ux{}}{\ux{}}
+ b_{5} \ux{} \sprod{\u}{\ux{}}^2 
+ b_{6} \u \sprod{\ux{3}}{\u} 
\\& & 
+ b_{7} \u \sprod{\ux{2}}{\ux{}} 
+ b_{8} \u \sprod{\ux{}}{\ux{}} \sprod{\u}{\ux{}} 
+ b_{9} \u \sprod{\ux{2}}{\u} \sprod{\u}{\ux{}} 
+ b_{10} \u \sprod{\ux{}}{\u}^3 
\end{eqnarray*}
where all coefficients $a_i,b_j$
are undetermined functions of $\sprod{\u}{\u},\sprod{\ux{}}{\ut}$.
The ansatz for the vector mKdV inverse flow 
with $\wt=-1,\wtau=3, \wu_U=1$ is:
\begin{eqnarray*}
\ux{t}\! &\! =\! &\! 
a_1 \ut \sprod{\u}{\ut}+ a_2 \u ,
\\ 
\uflow &\! =\! &\! 
b_{1} \ux{3} 
+ b_{2} \ux{2} \sprod{\u}{\ut}
+ b_{3} \ux{} \sprod{\ux{}}{\ut} 
+ b_{4} \ux{} \sprod{\u}{\ut}^2 
+ b_{5} \ux{} \sprod{\u}{\u} 
+ b_{6} \ut \sprod{\ux{}}{\ut}^2 
\\& & 
+ b_{7} \ut \sprod{\ux{}}{\ut} \sprod{\u}{\ut}^2 
+ b_{8} \ut \sprod{\ux{}}{\ut} \sprod{\u}{\u}
+ b_{9} \ut \sprod{\u}{\ut}^2 \sprod{\u}{\u} 
+ b_{10} \ut \sprod{\u}{\ut}^4 
\\& & 
+ b_{11} \ut \sprod{\u}{\ut} \sprod{\u}{\ux{}}
+ b_{12} \ut \sprod{\u}{\ux{2}}
+ b_{13} \ut \sprod{\ux{}}{\ux{}}
+ b_{14} \ut \sprod{\u}{\u}^2 
\\& & 
+ b_{15} \u \sprod{\ux{}}{\ut} \sprod{\u}{\ut} 
+ b_{16} \u \sprod{\u}{\ut} \sprod{\u}{\u}
+ b_{17} \u \sprod{\u}{\ut}^3 
+ b_{18} \u \sprod{\u}{\ux{}} 
\end{eqnarray*}
where all coefficients $a_i,b_j$
are undetermined functions of $\sprod{\ut}{\ut}$.
The ansatz for the vector NLS inverse flow is similar,
using a pair of vectors $U,V$, 
with $\wt=-2,\wtau=6,\wu_U=\wu_V=1$, 
and imposing the additional the filter 
$\wt=\wtau=0,\wu_U=1,\wu_V=-1$. 

\subsection{Computing Symmetry Conditions}

In the special case of evolutionary PDEs, 
all substitutions of $\phi^i_{t}$ and $\phi^i_{\tau}$ 
in the symmetry conditions 
$\phi^i_{[t,\tau]} := \D{\tau}{}\phi^i_{t} - \D{t}{}\phi^i_{\tau} =0$ 
are done only once because the \rhs/ of the PDEs and symmetries 
do not contain $t$- or $\tau$- derivatives. 
As a consequence,
the symmetry conditions are separately linear in the undetermined coefficients
of the PDEs and of the symmetries. 

The situation is different when the \lhs/ of 
the PDEs contain an $x$-derivative of $\phi^i$,
such as in the hyperbolic case $\phi^i_{tx}$. 
Then any $t$- or $x$-derivatives that appear in the \rhs/ of 
the PDEs and symmetries lead to (repeated) substitutions of 
$\phi^i_{tx}$ 
through the PDEs. 
The symmetry conditions become polynomially non-linear 
in the coefficients of the PDEs 
but stay linear in the coefficients of the symmetries. 
Due to this non-linearity, 
which is further amplified whenever the coefficients are 
functions of $t$-derivatives of $\phi^i$, 
the symmetry conditions can become extremely large.
The size of intermediate expressions can become exceedingly high
during the repeated
substitution and simplification process. 
To get over this memory hurdle, 
the linearity in the coefficients of the symmetry is exploited.
If the \rhs/ of the symmetry consists of $s$ terms 
with undetermined coefficients $b_1,\ldots,b_s$
then the full symmetry condition is computed via
\EQ
0 = \phi^i_{[t,\tau]} 
= \sum_{k=1}^s \left. \phi^i_{[t,\tau]} \right|_{\phi^j_{\tau}=g^j_{(k)}}
\end{equation} 
where $g^j_{(k)}$ denotes the $k$th term in the \rhs/ of 
$\phi^j_{\tau}=g^j$, 
i.e. all coefficients except $b_k$ are replaced by zero. 
This trade-off of computing memory for computing time is also justified 
by the fact that the time to simplify large expressions 
grows non-linearly with their size, 
at least in the computer algebra system {\sc Reduce} that is used.

\subsection{Solving the Symmetry Conditions}

Compared to the case of evolutionary PDEs,
the symmetry conditions for the general case of PDEs
with \lhs/ $\phi^i_{pt\,qx}$ 
are not purely algebraic but may involve 
ordinary and partial differential equations
and may be inhomogeneous and non-linear 
in the undetermined coefficients of the PDEs. 
The system of symmetry conditions is fed into 
the computer program {\sc Crack},
which was used to solve the algebraic symmetry conditions 
in the classification of evolutionary vector equations \cite{SokolovWolf} 
and coupled evolutionary scalar-vector systems \cite{TsuchidaWolf02}. 
Originally, {\sc Crack} has been developed as a solver for 
overdetermined PDE systems, and this functionality is essentially required
for solving the polynomially nonlinear differential system 
of symmetry conditions 
that occur in the classifications of vector hyperbolic equations 
in \secref{results}. 
Solutions of non-polynomial form, 
like square roots in the equations found in theorem~\ref{mkdvclass}, 
may appear and often are in reach of the {\sc Reduce} ODE solver 
that is employed by {\sc Crack} when appropriate. 

The strength of {\sc Crack} compared with other computer algebra packages 
for solving overdetermined systems lies in the flexibility of its approach, 
the possibility to run it in a fully automatic or interactive mode, 
and its support in handling especially large systems 
(safety, process control, 
heuristic algorithms to cut down the size of equations, 
ability to take advantage of the linearity of coefficient functions of 
the symmetries).
In addition, 
{\sc Crack} provides a verbose mode reporting all computational steps,
which allows in principle the solutions to be checked by (human) inspection
(in practise this is hard due to the high number of steps).

\section{ Sigma Model Interpretations }
\label{sigmamodels}

The wave equation $\scux{t}=0$ and SG equation $\scux{t}=\sin\scu$
are well-known to be intimately related to two of the simplest 
sigma models \cite{Pohlmeyer}, 
which are described by 
a geometrical wave equation for maps 
on $2$-dimensional Minkowski space $\R{1,1}$ into 
target spaces $S^1$ and $S^2$, respectively. 
In general a sigma model on $\R{1,1}$ is given by the nonlinear wave equation
\EQ
\label{sigmaeq}
0= \covder{t}{g}\der{x}\uvars{A}{}
= \der{t}\der{x}\uvars{A}{} 
+ \conx{A}{BC}(\uvars{}{}) \der{t}\uvars{B}{}\der{x}\uvars{C}{}
\doneEQ
for scalar functions $\uvars{A}{}(t,x)$
representing a map into some Riemannian target space $(M,\g{AB}{})$, 
$A=1,\ldots,\dim M$. 
Here $\covder{t}{g} = \der{t} +\conx{A}{BC}(\uvars{}{})\der{t}\uvars{B}{}$
is the pullback to $\R{1,1}$ of the (torsion-free) covariant derivative
determined by the metric $\g{AB}{}$
in local coordinates $\uvars{A}{}$ on the manifold $M$,
given in terms of the Christoffel symbol $\conx{A}{BC}$ of the metric. 
This wave equation has the geometrical action principle 
$S[\uvars{}{}] 
= \int \sprod{\der{t}\uvars{}{}}{\der{x}\uvars{}{}}_g \d t\d x$
where $\sprod{\cdot}{\cdot}_g$ denotes the inner product 
in $T(M)$ given by the metric $\g{AB}{}(\uvars{}{})$. 
Another geometrical meaning for this wave equation arises
if we view $\uvars{A}{}(t,x)$ for fixed $t$ as a curve 
embedded into the manifold $M$.
The sigma model thereby describes a flow of this curve,
whose arclength is preserved due to the conservation law
\EQ
\label{sigmaDtconslaw}
\der{t}( \der{x}\uvars{A}{}\der{x}\uvars{B}{} \g{AB}{}(u) ) =0 . 
\doneEQ
Because of the reflection symmetry $t \leftrightarrow x$ in the model, 
there is also a conservation law
\EQ
\label{sigmaDxconslaw}
\der{x}( \der{t}\uvars{A}{}\der{t}\uvars{B}{} \g{AB}{}(u) ) =0 . 
\doneEQ
These conservation laws are connected with 
invariance of the model under conformal scaling transformations 
$t \rightarrow t':= \alpha(t)$, 
$x \rightarrow x':= \beta(x)$. 
This freedom can be fixed in a natural manner by putting
$\alpha'(t)^2 = \sprod{\der{t}\uvars{}{}}{\der{t}\uvars{}{}}_g$
and 
$\beta'(x)^2 = \sprod{\der{x}\uvars{}{}}{\der{x}\uvars{}{}}_g$.
Finally, note that in general the sigma model equation \eqref{sigmaeq}
is $\wu$-homogeneous with scaling weight $\wu=0$ on $\uvars{A}{}$
while $t$ and $x$ have opposite weights of $\wt=-1$ and $+1$,
in the conventions of \secref{results}. 

$O(N)$-invariant vector wave equations 
arise from sigma models in various ways
and for certain types of target spaces. 
One direct way is by considering target spaces 
embedded as 
an $O(N)$ symmetric hypersurface 
in Euclidean space $\R{N+1}$ or Minkowski space $\R{N,1}$. 
Vector coordinates for the hypersurface given by its embedding
lead to natural vector generalizations 
of the variant scalar wave equation 
$\scux{t} = -\scu(1-\scu^2)^{-1} \scut\scux{}$
which is the sigma model for $S^1 \subset \R{2}$. 
A second way involves introducing a flat Cartan connection and frame bundle 
on homogeneous target spaces with the tangent space structure 
$\alg{g}/\alg{h} \simeq \R{N+1}$ 
based on pairs of Lie algebras $\alg{g} \supset \alg{h}$. 
For a suitable geometric choice of frame, the connection components 
yield natural vector generalizations of the variant SG equation 
$\scux{t} = \scu\sqrt{1-\scut^2}$
which comes from the sigma model for 
$S^2 \simeq SO(3)/SO(2)$. 

In \secref{SGclassderivation} we show that 
the vector wave equation \eqref{vectorSGeq} in theorem~\ref{sgclass} 
arises from the nonlinear sigma model for 
an $O(N)$ symmetric paraboloid hypersurface in $\R{N+1}$.
In \secref{invflowclassderivation} we derive
the first two hyperbolic vector equations in theorem~\ref{mkdvclass}
from the flat Cartan connection of gauged sigma models 
viewed as a flow of curves
on homogeneous spaces (Klein geometries) 
associated with the Lie algebras
$\alg{so}(N+1) \subset \alg{so}(N+2),\alg{su}(N+1)$. 

\subsection{ $N$-sphere and $N$-dimensional paraboloid sigma models }
\label{SGclassderivation}

Consider firstly the $N$-sphere target space $S^N$ 
as an embedded hypersurface in $\R{N+1}$. 
We describe $S^N$ by the $N+1$-component unit vector 
$\uvars{i}{}$, $i=1,\ldots,N+1$, constrained to satisfy 
$1= \uvars{i}{}\uvars{j}{} \id{ij}{}$. 
If we resolve the constraint by expressing
$\uvars{N+1}{} = \pm \sqrt{ 1-\uvars{A}{}\uvars{B}{}\id{AB}{} }$
with $\uvars{i}{} = (\uvars{A}{},\uvars{N+1}{})$, $A=1,\ldots,N$, 
then $\uvars{A}{}$ represents a set of coordinates on $S^N$. 
In particular the upper or lower hemisphere of $S^N$ is mapped
by $\uvars{A}{}$ to the open unit ball in $\R{N}$. 
In these coordinates, the Riemannian metric on $S^N$ 
as induced by the Euclidean metric $\id{ij}{}$ on $\R{N+1}$
is given by 
\EQ
\g{AB}{}(\uvars{}{})
= \id{AB}{} +(1- \uvars{C}{}\uvars{}{C})^{-1} \uvars{}{A}\uvars{}{B}
\doneEQ
with the inverse metric
$\g{}{AB} = \id{}{AB} -\uvars{A}{}\uvars{B}{}$,
where $\uvars{}{A} := \id{AB}{} \uvars{B}{}$. 
The Christoffel symbol and covariant derivative on $S^N$ 
are readily found from the condition $\covder{C}{g} \g{}{AB}(u) =0$,
giving 
\EQ
\covder{C}{g} = \deru{C} + \uvars{A}{} \g{BC}{}(\uvars{}{}) ,\quad
\conx{A}{BC}(\uvars{}{}) 
= \uvars{A}{}( \id{BC}{} +(1- \uvars{D}{}\uvars{}{D})^{-1} 
\uvars{}{B}\uvars{}{C} ) . 
\doneEQ
Hence the curvature is given by 
\EQ
[\covder{A}{g},\covder{B}{g}] := \curv{ABC}{D} 
= 2 \g{C[A}{}(\uvars{}{}) \id{B]}{D} \scurv , \quad
\scurv = 1
\doneEQ
where $\scurv$ denotes the scalar curvature. 
Note this indicates $S^N$ is an $O(N)$ symmetric space of
(positive) constant curvature. 

The sigma model equation \eqref{sigmaeq} for the target space $S^N$
thus has the form 
\EQ
\label{nspheresigmaeq}
0 = 
\der{t}\der{x}\uvars{A}{} 
+ \uvars{A}{}( \der{t}\uvars{B}{}\der{x}\uvars{}{B}
+ (1-\uvars{B}{}\uvars{}{B})^{-1} \uvars{}{C} \uvars{}{D} 
\der{t}\uvars{C}{} \der{x}\uvars{D}{} )
\doneEQ
which is an $O(N)$-invariant hyperbolic vector equation 
for $\uvars{A}{}(t,x)$. 
The scalar case $N=1$ is seen to reduce to 
$0= \scux{t} + \scu(1-\scu^2)^{-1} \scut\scux{}$. 
Integrability of this sigma model for all $N\geq 1$ 
is a well-established result 
via a Lax pair associated with $S^N = SO(N+1)/SO(N)$
as a Riemannian symmetric space 
\cite{ZacharovMikhailov,Pohlmeyer},
and also more recently via a bi-Hamiltonian structure 
(higher-symmetry recursion operator)
derived from a moving frame formulation of $S^N$ 
as a constant curvature space 
\cite{SandersWang-curveflows,AncoWang}. 
Its \hsymms/ were first found in 
\cite{Pohlmeyer}
using a B\"acklund transformation technique
which relies on conformally scaling 
$t\rightarrow t'$, $x\rightarrow x'$ 
through the conservation laws \eqrefs{sigmaDtconslaw}{sigmaDxconslaw}. 
The hierarchy of \hsymms/ is homogeneous
and has positive scaling weight $\wtau=3,5,\ldots$, 
with respect to the scaling group 
$(x',t',\uvars{A}{}) \rightarrow 
(e^{\epsilon}x',e^{-\epsilon}t',\uvars{A}{})$, 
$\epsilon \in \R{}$. 
But when expressed in the original variables $t,x$, 
the \hsymms/ are all homogeneous of weight $\wtau=0$. 
For this reason the vector wave equation \eqref{nspheresigmaeq}
does not appear in our classifications. 

To account for the vector wave equation \eqref{vectorSGeq}
and its \hsymm/ \eqref{vectorSGsymm} 
with SG scaling weight
in theorem~\ref{sgclass}, 
we look for a target space with an appropriate Riemannian metric. 
Working backward from the nonlinear terms in the equation
\EQ
\label{nparaboloidsigmaeq}
0 = 
\der{t}\der{x}\uvars{A}{} 
+ \uvars{A}{} (1+\uvars{C}{}\uvars{}{C})^{-1} 
\der{t}\uvars{B}{}\der{x}\uvars{}{B}
\doneEQ
we deduce the Christoffel symbol 
\EQ
\conx{A}{BC}(\uvars{}{}) 
= \uvars{A}{} \id{BC}{} (1+ \uvars{D}{}\uvars{}{D})^{-1} 
\doneEQ
and hence 
\EQ
\covder{C}{g} 
= \deru{C} + \uvars{A}{} \id{BC}{} (1+ \uvars{D}{}\uvars{}{D})^{-1}
\doneEQ
is the covariant derivative. 
 From the condition $\covder{C}{g} \g{}{AB}(u) =0$,
it is straightforward to derive 
$\g{}{AB} 
= \id{}{AB} +(1+ \uvars{C}{}\uvars{}{C})^{-1} \uvars{A}{}\uvars{B}{}$
which then gives the metric
\EQ
\g{AB}{}(\uvars{}{})
= \id{AB}{} -\uvars{}{A}\uvars{}{B} . 
\doneEQ
Finally we are able to identify this metric 
as the induced Riemannian metric
on the $N$-dimensional target space given by the paraboloid hypersurface
\EQ
\uvars{N+1}{} = \frac{1}{2} \uvars{A}{}\uvars{B}{}\id{AB}{} 
\quad\eqtext{ in $\R{N+1}$.  }
\doneEQ
This target space is an $O(N)$ symmetric Riemannian manifold,
whose scalar curvature is non-constant
\EQ
\scurv = (1 + 2\uvars{N+1}{})^{-1} ,\quad 
\curv{ABC}{D}
= 2 \id{C[A}{} \g{B]}{D}(\uvars{}{}) \scurv
\doneEQ
where $\g{A}{B} := \id{AC}{} \g{}{CB}$. 
(Curiously, 
the metric tensor $\g{A}{B}(\uvars{}{})$ 
on the $N$-dimensional paraboloid 
is the same as the inverse metric tensor 
$\ginv{B}{A}(\uvars{}{}) := \ginv{BC}{}(\uvars{}{}) \id{CA}{}$ 
on the $N$-sphere.)

We note there is a Lorentzian version of these sigma models 
\eqrefs{nspheresigmaeq}{nparaboloidsigmaeq}
obtained via replacing the Euclidean space $\R{N+1}$ 
by Minkowski space $\R{N,1}$. 
Correspondingly, $S^N$ is replaced by its Lorentzian counterpart
given by the $N$-dimensional hyperboloid space 
$\uvars{N+1}{} = \pm \sqrt{ 1+\uvars{A}{}\uvars{B}{}\id{AB}{} }$
which comes from the spacelike hypersurface
$1= \uvars{i}{}\uvars{j}{} \flat{ij}{}$ in $\R{N,1}$, 
where $\flat{ij}{}$ is the Minkowski metric
and $\uvars{i}{} = (\uvars{A}{},\uvars{N+1}{})$. 
Its induced Riemannian metric is 
$\g{AB}{}(\uvars{}{})
= \id{AB}{} +(1+ \uvars{C}{}\uvars{}{C})^{-1} \uvars{}{A}\uvars{}{B}$,
with constant (negative) scalar curvature $\scurv=-1$. 
In contrast, the $N$-dimensional paraboloid space 
$\uvars{N+1}{} = \frac{1}{2} \uvars{A}{}\uvars{B}{}\id{AB}{}$
is its own Lorentzian counterpart,
but with the metric replaced by 
$\g{AB}{}(\uvars{}{})
= \id{AB}{} +\uvars{}{A}\uvars{}{B}$
as induced by the Minkowski metric $\flat{ij}{}$,
and with non-constant scalar curvature
$\scurv = (1 - 2\uvars{N+1}{})^{-1}$. 
Note in the Lorentzian case this paraboloid hypersurface
has a null tangent direction at points $2\uvars{N+1}{}=1$ in $\R{N,1}$
where both the metric and curvature become singular. 

Thus the nonlinear sigma model \eqref{nparaboloidsigmaeq} 
for the $N$-dimensional paraboloid target space 
in the Euclidean and Lorentzian cases 
reproduces the vector wave equation \eqref{vectorSGeq}
with the vector $\u$ identified as the local coordinates $\uvars{A}{}$. 
Integrability of this model does not seem to be known in the literature,
so its \hsymm/ \eqref{vectorSGsymm} stated in theorem~\ref{sgclass}
is a new result. 
Unlike the $S^N$ model,
the existence of a hierarchy of \hsymms/ is an open question,
apart from the scalar case $N=1$ of this model \eqref{nparaboloidsigmaeq} 
which gives the same integrable $S^1$ sigma model 
as the scalar reduction of the $S^N$ model \eqref{nspheresigmaeq}. 
As noted in the theorem, for $N\ge 1$
we checked that there are \hsymms/ of all weights $\wtau=4,5,6,7$. 
This strongly suggests the model \eqref{nparaboloidsigmaeq} 
is indeed integrable. 
A proof will require exhibiting e.g. 
a Lax pair, a bi-Hamiltonian structure, or a symmetry-recursion operator.

\subsection{ Homogeneous (gauged) sigma models and curve flows }
\label{invflowclassderivation}

We now turn to the homogeneous target spaces 
$M=G/H$ based on Lie groups $H \subset G$
chosen to be semi-simple and compact,
with the structure of a symmetric space. 
This structure is expressed by the Lie algebra decomposition
$\alg{g} = \alg{h} \oplus \alg{\quot}$
such that $[\alg{h},\alg{\quot}] \subseteq \alg{\quot}$
and $[\alg{\quot},\alg{\quot}] \subseteq \alg{h}$,
where $\alg{h},\alg{g}$ denote the respective Lie algebra of $H,G$,
and $\alg{\quot}$ is a vector space 
canonically identified (through left-invariant vector fields) 
with the tangent space ${TM}$ at any point in $M$. 
As is well-known, these target spaces carry the structure of 
a symmetric Riemannian manifold whose metric tensor is given by 
the Cartan-Killing inner product $\sprod{\cdot}{\cdot}$
on $TG \simeq \alg{g}$ restricted to ${TM} \simeq \alg{p}$. 
This metric and its curvature tensor are both $G$-invariant
and covariantly constant. 
A few main examples are 
$SO(k+1)/SO(k)\simeq S^k$, 
$SU(k+1)/U(k)\simeq {\C}P^k$, 
$SU(k)/SO(k)$
(see \cite{Helgason} for a complete list and properties). 
The sigma model \eqref{sigmaeq} 
for all such target spaces is a gauged chiral equation \cite{EichenherrForger}
\EQ
\label{gaugedsigmaeq}
0= \der{t}(\uvars{-1}{}\der{x}\uvars{}{})\upindex{A}
= \covder{t}{\alg{h}} \der{x}\uvars{A}{}
\doneEQ
with $\uvars{A}{}(t,x)$ given by gauge equivalence classes of 
$G$-valued functions under the gauge group $H \subset G$,
where $\covder{}{\alg{h}}$ denotes the gauge covariant derivative. 
Viewed as a flow of curves in $M =G/H$, 
this model has a natural frame bundle formulation 
employing a flat Cartan connection 
(identified with the Maurer-Cartan form on $G$)
whose components will satisfy 
an $O(N)$-invariant nonlinear vector wave equation
\cite{DauriaReggeScuito}. 
Details of Cartan connections and frame bundles 
for homogeneous spaces are presented in \cite{Sharpe}.

Firstly, 
we consider ${TM} \simeq \R{N+1} = \alg{\quot}$ 
as given by 
$\alg{g} = \alg{so}(N+2)$ and $\alg{h} = \alg{so}(N+1)$. 
Recall, $\alg{so}(k)$ is a real vector space of dimension $\frac{k(k-1)}{2}$ 
isomorphic to the Lie algebra of $k\times k$ skew-symmetric matrices. 
So ${TM}= \alg{so}(N+2)/\alg{so}(N+1)$ 
is isomorphic to $\alg{\quot} = \R{N+1}$,
as described by the following canonical decomposition 
\EQ
\label{sphereliealg}
\begin{pmatrix}
0 & \quot \\ -\quot^\T & \matrh 
\end{pmatrix}
\in \alg {so}(N+2) ,\quad
\matrh \in \alg {so}(N+1) ,\quad
\quot \in \R{N+1}
\doneEQ
parameterized by the $N+1$-component vector $\quotvec{A}$, 
$A=1,\ldots,N+1$. 
Note the Cartan-Killing inner product on $\alg{g}$ is 
(up to a normalization factor)
the trace of the product of $\alg{so}(N+2)$ matrices \eqref{sphereliealg}
and hence reduces for $\matrh=0$ to the ordinary dot product $\id{AB}{}$
on vectors $\quotvec{A}$. 
On the target space 
$M = SO(N+2)/SO(N+1)$ 
we introduce a flat Cartan connection $\algconx{g}{A}$
which is $\alg{so}(N+2)$-valued 
with the decomposition 
$\algconx{g}{A} = \algconx{h}{A} + \algconx{\quot}{A}$. 
As a consequence of the Lie algebra structure \eqref{sphereliealg},
$\algconx{h}{A}$ provides an $\alg{so}(N+1)$-valued connection 
to which is associated an $\R{N+1}$-valued orthonormal frame $\e{A}{}$ 
for the tangent space 
${TM} \simeq \alg{so}(N+2)/\alg{so}(N+1) \simeq \R{N+1}$. 
More precisely, the underlying (gauge) structure group here is $SO(N+1)$,
namely the matrix Lie group associated with $\alg{h}=\alg{so}(N+1)$. 
Up to the gauge freedom of $SO(N+1)$ rotations, 
there is a canonical soldering relation $\algconx{\quot}{A} = \e{}{A}$
between the frame and the reduced connection,
where $\e{}{A}$ is the coframe satisfying 
$\sprod{\e{}{A}}{\e{B}{}}_\alg{\quot} = \id{A}{B}$. 
This soldering can be shown to encode 
a nonzero Riemannian curvature of the metric on $M$,
determined by the coframe $\e{}{A}$ and the connection $\algconx{h}{A}$. 

To derive an $O(N)$-invariant vector wave equation from 
the gauged sigma model \eqref{gaugedsigmaeq} 
for this target space $M= SO(N+2)/SO(N+1)$, 
we first write 
$\algtconx{h}{} := \algconx{h}{A}{} \der{t}\uvars{A}{}$, 
$\algxconx{h}{} := \algconx{h}{A}{} \der{x}\uvars{A}{}$, 
$\te{} := \e{}{A} \der{t}\uvars{A}{}$, 
$\xe{} := \e{}{A} \der{x}\uvars{A}{}$,
in terms of $\uvars{A}{}(t,x)$. 
These obey 
\EQ
\label{spheretorsion}
0= \der{t} \xe{} + \algtconx{h}{} \xe{} 
= \der{x} \te{} + \algxconx{h}{} \te{} 
\doneEQ
as a consequence of the chiral equation \eqref{gaugedsigmaeq}
and the structure equation of the soldering frame 
$\algconx{\quot}{A} = \e{}{A}$. 
Interpreted geometrically,
the second equality is the Cartan structure equation for
vanishing torsion of the connection $\algconx{g}{A}$. 
The Cartan structure equation for the curvature of this flat connection 
gives
\EQ
\label{spherecurv}
\der{x} \algtconx{h}{} - \der{t} \algxconx{h}{}
+ [\algxconx{h}{},\algtconx{h}{}] 
= [\algtconx{p}{},\algxconx{p}{}]
= \te{} \wedge \xe{}{}
\doneEQ
where $\wedge$ denotes the antisymmetric tensor product. 
Note from equation \eqref{spheretorsion}
we recover the conservation laws \eqrefs{sigmaDtconslaw}{sigmaDxconslaw}
for this sigma model, 
$0=\der{t}\sprod{\xe{}}{\xe{}}_\alg{\quot} 
=\der{x}\sprod{\te{}}{\te{}}_\alg{\quot}$. 
By conformally scaling $t,x$,
we normalize 
$\sprod{\xe{}}{\xe{}}_\alg{\quot}=\sprod{\te{}}{\te{}}_\alg{\quot} =1$.
Then we use the inherent gauge freedom on $\e{}{A}$
using the rotation group $SO(N+1)$ to put
\EQ\label{adapted}
\xe{N+1} =1 ,\quad 
\xe{i} =0 , 
\doneEQ
without loss of generality. 
Thus one of the conservation laws is satisfied identically; 
the other one determines 
\EQ
\label{sphereconslawid}
\te{N+1} = \pm \sqrt{ 1- \te{i} \te{j} \id{ij}{} } . 
\doneEQ
In the index notation here, 
$\xe{} =(\xe{i},\xe{N+1})$ and $\te{} =(\te{i},\te{N+1})$, 
$i=1,\ldots,N$, 
are $(N+1\times 1)$ column matrices comprising 
the frame components of, respectively, 
the tangent vector $\der{x}\uvars{A}{}$ along the curve $\uvars{A}{}(t,x)$ 
and the direction vector $\der{t}\uvars{A}{}$ of the flow
on $M= SO(N+2)/SO(N+1)$. 
The condition \eqref{adapted} geometrically means 
$\e{A}{}$ restricted to the curve $\uvars{A}{}(t,x)$
is an adapted moving frame,
consisting of $N$ normal vectors and one tangential vector 
(so $x$ has the geometrical meaning of arclength). 
There now remains $SO(N)$ gauge freedom 
that preserves the adapted form \eqref{adapted} of $\xe{}$, 
which is described by rotations in the normal space along the curve. 
We use this freedom to fix the connection $\algxconx{h}{}$
so it corresponds geometrically to specializing 
the adapted moving frame to be parallel (cf. \cite{Bishop})
with respect to the curve $\uvars{A}{}(t,x)$ as follows. 
Write 
$\algxconx{h}{}
= (\algxconx{h}{ij},\algxconx{h}{iN+1})$
and 
$\algtconx{h}{}
= (\algtconx{h}{ij},\algtconx{h}{iN+1})$
for the components of the ($N+1\times N+1$) connection matrices.
The curvature equation \eqref{spherecurv} shows that
the pair of matrices $\algxconx{h}{ij},\algtconx{h}{ij}$ constitutes
a flat $\alg{so}(N)$-valued connection $1$-form, 
and consequently along the curve 
we put $\algxconx{h}{ij}=0$
by the available gauge freedom. 
This leads to $\algtconx{h}{ij}=0$ without loss of generality. 
Geometrically speaking,
$\algxconx{h}{}$ then describes the derivative of $\e{A}{}$
under infinitesimal displacement along the curve $\uvars{A}{}(t,x)$
such that the normal vectors in the frame 
have a purely tangential derivative.  

Now, the torsion equation \eqref{spheretorsion} can be used to find
\EQs
&&
\algtconx{h}{iN+1}{} =0 , 
\\
&&
\algxconx{h}{iN+1}{} 
= (\te{N+1})^{-1} \der{x} \te{i} , 
\label{spherexconx}
\doneEQs
while from the curvature equation \eqref{spherecurv}
we have 
\EQ
\te{i} 
= -\der{t} \algxconx{h}{iN+1}{} . 
\label{spheretconx}
\doneEQ
Thus the reduced Cartan equations \eqrefs{spherexconx}{spheretconx}
together with the relation \eqref{sphereconslawid}
yield a system of coupled vector equations 
$\der{x} \te{i} = 
\pm ( 1- \te{i} \te{j} \id{ij}{} )^{1/2} \algxconx{h}{iN+1}{}$
and 
$\der{t} \algxconx{h}{iN+1}{} = -\te{i}$. 
Viewing $\algxconx{h}{iN+1}{}$ 
as an $N$-component vector $U(t,x)$,
and eliminating $\te{i}$ from these equations,
we obtain precisely the $O(N)$-invariant nonlinear vector wave equation 
\eqref{invmkdvIflow}
in theorem~\ref{mkdvclass}. 
As shown by the theorem,
this vector wave equation possesses a \hsymm/ 
given by the $+1$ flow in the vector mKdV hierarchy \eqref{mkdveqI}. 

Integrability of the nonlinear vector wave equation \eqref{invmkdvIflow}
was first shown via a Lax pair 
\cite{EichenherrPohlmeyer,Bakis}.
Related derivations of it using both a zero-curvature
and a Lie-algebraic formulation of the Lax pair were obtained in 
\cite{PohlmeyerRehren,DauriaReggeScuito,Wang}. 
Our derivation here is more intrinsically geometrical,
based on the approach in 
\cite{Mari-BefaSandersWang,SandersWang-curveflows,AncoWang}
relying on use of 
a parallel moving frame and a corresponding adapted connection 
\EQ
\xe{}=(0,\ldots,0,1) \in \R{N+1},\quad
\algxconx{h}{} = 
\begin{pmatrix}
0 & \u \\ -\u^\T & \zero
\end{pmatrix} 
\in \alg{so}(N+1),\quad
\u \in \R{N}
\doneEQ
for a curve $\uvars{A}{}(t,x)$
associated to the sigma model \eqref{gaugedsigmaeq}
as interpreted as a flow of curves 
on the Riemannian space 
$SO(N+2)/SO(N+1)\simeq S^{N+1}$.
The connection is related to the frame by the Frenet-Serret formula
$\covder{x}{\alg{h}} \e{}{A} = -\algxconx{h}{} \e{}{A}$. 
In particular, in this frame 
the connection components $\u$ geometrically represent 
differential invariants of the curve. 

The second $O(N)$-invariant vector wave equation \eqref{invmkdvIIflow}
in theorem~\ref{mkdvclass}
has a similar derivation,
starting from the target space $M=SU(N+1)/SO(N+1)$.
One notable difference in this sigma model is that 
the target space has dimension greater than $N+1$
and hence a reduction condition must be imposed
on the Cartan connection and its associated soldering frame,
allowing them to be expressed in terms of a $N+1$-component vector.
The \hsymm/ structure 
for this vector wave equation obtained by theorem~\ref{mkdvclass},
identified as the $+1$ flow in the vector mKdV hierarchy \eqref{mkdveqII}, 
is a new result. 
Of course, a proof of its integrability will require e.g. 
checking that it admits the symmetry-recursion operator of 
the hierarchy. 
The same remarks apply to the other two 
vector wave equations \eqrefs{invnlsIflow}{invnlsIIflow}
in theorem~\ref{mkdvclass}. 

To outline the derivation of equation \eqref{invmkdvIIflow}, 
recall $\alg{su}(k)$ is a complex vector space 
isomorphic to the Lie algebra of $k\times k$ skew-hermitian matrices.
The real and imaginary parts of these matrices belong to 
the real vector space $\alg{so}(k)$ of skew-symmetric matrices
and the real vector space $\alg{s}(k) \simeq \alg{su}(k)/\alg{so}(k)$
defined by $k\times k$ symmetric trace-free matrices. 
Hence $\alg{g}= \alg{su}(N+1)$ has the decomposition 
$\alg{g} =\alg{h} +\i \alg{p}$ 
where $\alg{h}=\alg{so}(N+1)$
and $\alg{p} =\alg{s}(N+1)$. 
There is a natural subset of matrices in $\alg{su}(N+1)$
given by the special form 
\EQ
\label{reduced}
\frac{1}{N+1} \quot\cdot\quot\ \I - \quot^\T\quot \in \alg{s}(N+1) ,\quad 
\quot \in \R{N+1} 
\doneEQ
which is parametrized by the arbitrary $N+1$-component vector $\quotvec{A}$.
This set of symmetric trace-free matrices \eqref{reduced},
hereafter denoted $\alg{s}_\quot(N+1)$,
provides a canonical reduction of $\alg{su}(N+1)/\alg{so}(N+1)$ to $\R{N+1}$
as a $N+1$-dimensional manifold. 
The restriction of the Cartan-Killing metric of $\alg{su}(N+1)$ 
to $\alg{s}_\quot(N+1)$ yields a Riemannian metric 
$\sprod{\cdot}{\cdot}_{\alg{s}_\quot}$ 
proportional to the trace of the product of matrices \eqref{reduced}.
In particular, the norm of a matrix \eqref{reduced} in this metric
(with a suitable normalization)
is simply $\quot\cdot\quot$. 
(Note the nonlinearity of the matrix norm in terms of $\quot$
is due to the lack of a vector space structure for $\alg{s}_\quot(N+1)$.)
We now introduce an $\alg{su}(N+1)$-valued flat Cartan connection 
$\algconx{g}{A} = \algconx{h}{A} + \i \algconx{s}{A}$,
with $\algconx{h}{A}$ defining an $\alg{so}(N+1)$-valued connection 
to which is associated 
an $\alg{s}_\quot(N+1)$-valued soldering frame $\e{A}{}$ 
and a corresponding coframe $\e{}{A}=\algconx{s}{A}$
such that $\sprod{\e{}{A}}{\e{B}{}}_{\alg{s}_\quot} = \id{A}{B}$. 
There is $SO(N+1)$ gauge freedom inherent in $\e{}{A}$ and $\algconx{h}{A}$.
This structure is a reduction of the full soldering frame 
and flat Cartan connection attached to the target space
$M=SU(N+1)/SO(N+1)$.

To proceed we impose this reduction on 
the target space in the gauged sigma model \eqref{gaugedsigmaeq}
for $\uvars{A}{}(t,x)$. 
Thus, along the tangent vector $\der{x}\uvars{A}{}$ 
and the flow direction vector $\der{t}\uvars{A}{}$
for the curve $\uvars{A}{}(t,x)$,
we have the frame matrices $\xe{},\te{}$
and the connection matrices $\algxconx{h}{},\algtconx{h}{}$
that obey the torsion and curvature equations
\EQs
&&
0= \der{t} \xe{} + \algtconx{h}{} \xe{} 
= \der{x} \te{} + \algxconx{h}{} \te{} , 
\label{reducedtors}\\
&&
\der{x} \algtconx{h}{} - \der{t} \algxconx{h}{}
+ [\algxconx{h}{},\algtconx{h}{}] 
= [\algxconx{s}{},\algtconx{s}{}] 
= [\xe{},\te{}{}] .
\label{reducedcurv}
\doneEQs
Using the sigma model conservation laws 
\eqrefs{sigmaDtconslaw}{sigmaDxconslaw}, 
we put 
$\sprod{\xe{}}{\xe{}}_{\alg{s}_\quot}
=\sprod{\te{}}{\te{}}_{\alg{s}_\quot} 
=1$.
We further fix $\xe{},\algxconx{h}{}$
(via the $SO(N+1)$ gauge freedom)
so they represent a parallel moving frame and an adapted connection 
along the curve $\uvars{A}{}(t,x)$. 
Namely, in matrix notation,
\EQ
\xe{} = 
\frac{1}{N+1} \I - 
\begin{pmatrix}
1 & 0 \\ 0 & \zero
\end{pmatrix}
\in \alg{s}_\quot(N+1) ,\quad
\algxconx{h}{} = 
\begin{pmatrix}
0 & \u\\ -\u^\T & \zero
\end{pmatrix}
\in \alg{so}(N+1) ,\quad
\u \in \R{N}
\doneEQ
where $\zero \in \alg{so}(N)$,
and also 
\EQ
\te{} = 
\frac{1}{N+1} \I - 
\begin{pmatrix}
v & \v\\ \v^\T & v^{-1} \v^\T \v
\end{pmatrix}
\in \alg{s}_\quot(N+1) ,\quad
(\v,v) \in \R{N+1}
\doneEQ
where
\EQ
\label{vVrel}
v^2 +\v\cdot \v = v .
\doneEQ
Note $\xe{},\te{}$ are of the form \eqref{reduced}
with respective parameterizations chosen in vector index notation as 
$1=\quotvec{N+1}$, $0=\quotvec{i}$, 
and $\sqrt{v}=\quotvec{N+1}$, $(\sqrt{v})^{-1}\vvar{i}=\quotvec{i}$;
the norm of $\te{}$ is 
$1=(\quotvec{N+1})^2 + \quotvec{i}\quotvec{j}\id{ij}{}
= v +v^{-1} \vvar{i}\vvar{j}\id{ij}{}$. 
Now, 
from the $\alg{so}(N)$ part of the curvature equation \eqref{reducedcurv}
together with the first half of the torsion equation \eqref{reducedtors},
we find 
$\algtconx{h}{} = 
\begin{pmatrix}
0 & 0 \\ 0 & \zero
\end{pmatrix}$.
Then the remaining parts of equations \eqrefs{reducedtors}{reducedcurv}
reduce to
\EQs
&&
\der{x}\v = v\u - v^{-1} {\u\cdot\v}\ \v ,\quad
\der{t} \u = -\v ,
\label{reducedtorscurv}\\
&& 
\der{x}v = -2 {\u\cdot\v} ,\quad
\der{x}( v^{-1}\v^T \v) = \v^\T\u + \u^\T\v .
\label{reducedids}
\doneEQs
We see equations \eqrefs{reducedtorscurv}{vVrel}
are equivalent to the $O(N)$-invariant nonlinear vector wave equation
\eqref{invmkdvIIflow}, 
while the last equation \eqref{reducedids} is found to hold as an identity. 

\subsection{ Other models and concluding remarks }

The sigma models connected with 
the two $O(N)$-invariant nonlinear vector wave equations
\eqrefs{invmkdvIflow}{invmkdvIIflow} 
in theorem~\ref{mkdvclass}
exhaust all the types of real symmetric Riemannian spaces $G/SO(N+1)$. 
The other two vector wave equations
\eqrefs{invnlsIflow}{invnlsIIflow}, 
which in contrast are complex-valued and $U(N)$-invariant,  
might then be expected to arise from 
hermitian symmetric spaces $G/U(N+1)$. 
However, sigma models based on the latter type of target space 
naturally take the form of coupled scalar-vector wave equations,
and hence degenerate cases would have to be investigated in order 
to account for these purely vectorial wave equations
\eqrefs{invnlsIflow}{invnlsIIflow}. 

Interestingly, 
the last vector wave equation \eqref{invnlsIIflow} 
given in theorem~\ref{mkdvclass}
stands out from the others in one respect; 
it is neither well-defined in the scalar case $N=1$
nor in the real-valued vector case 
(precisely when the expression \eqref{invnlsIIcoeffs} vanishes). 
Thus it exhibits an intrinsic complex-valued vectorial nature. 
Some insight on this situation comes from the NLS \hsymm/ 
\eqref{invnlsIIsymm} 
admitted by this vector wave equation. 
In \cite{KulishSklyanin}
the corresponding vector NLS hierarchy \eqref{nlseqII}
is shown to originate from
the matrix NLS equation 
$\i\der{t}\matru = \nder{x}{2}\matru + \matru\matru^\dagger\matru$
where the matrix $\matru=\uvars{A}{}\gamma_A$ is chosen to belong to 
the span of the gamma matrices $\gamma_A$, $A=1,\ldots,N$,
defined to satisfy the Clifford algebra relations 
$\gamma_A \gamma_B + \gamma_B \gamma_A =2\id{AB}{} \I$.
This observation suggests the vector wave equation \eqref{invnlsIIflow}
may arise in a similar fashion from a suitable matrix type of sigma model.

Finally, in connection with these remarks,
we mention the deep work 
\cite{alg1,alg2,alg3}
of Svinolupov and Sokolov
on the construction of integrable vector and matrix equations of
hyperbolic and evolutionary type,
related to special kinds of nonassociative algebras.

\section*{Acknowledgements}

One author (TW) wants to thank Winfried Neun for discussions, and the
Konrad Zuse Institute Berlin for support through its fellowship program. 
Takayuki Tsuchida is thanked for comments on derivative-NLS $-1$ flows. 
Computations were performed partially on computer clusters
of the SHARCNET consortium (www.sharcnet.com).

\label{lastpage}

\end{document}